\documentclass{article}

\usepackage{epsfig}
\def\ligne#1{\hbox to \hsize{#1}}
\def\PlacerEn#1 #2 #3 {\rlap{\kern#1\raise#2\hbox{#3}}}
\def\cqfd{\hbox{\kern 2pt\vrule height 6pt depth 2pt width 8pt\kern 1pt}}
\def\leurre{\noindent\leftskip0pt\small\baselineskip 10pt}

\newtheorem{thm}{Theorem}
\newtheorem{lem}{Lemma}

\newtheorem{cor}{Corollary}
\newtheorem{fig}{Figure}
\newtheorem{tab}{Table}

\font\bfix=cmbx9

\font\rmix=cmr9 \font\mmix=cmmi9 \font\symix=cmsy9
\def\mathix{\textfont0=\rmix \textfont1=\mmix \textfont2=\symix}

\def\grostrait{\ligne{\vrule height 1pt depth 1pt width \hsize}}
\def\demitrait{\ligne{\vrule height 0.5pt depth 0.5pt width \hsize}}

\begin{document}
\begin{center}
{\bf \Large About the embedding of one dimensional cellular automata into 
hyperbolic cellular automata}
\vskip 5pt
Maurice {\sc Margenstern}
\vskip 2pt
Universit\'e Paul Verlaine $-$ Metz, IUT de Metz,\\
LITA EA 3097, UFR MIM,\\
Campus du Saulcy,\\
57045 METZ C\'edex 1, FRANCE\\
{\it e-mail}: {\tt margens@univ-metz.fr}\\
{\it Web page}: {\tt http://www.lita.sciences.univ-metz.fr/\~{}margens}
\end{center}
\vskip 5pt
{\parindent 0pt\leftskip 20pt\rightskip 20pt
{\bf Abstract} $-$ In this paper, we look at two ways to implement one dimensional
cellular automata into hyperbolic cellular automata in three contexts: the pentagrid, the
heptagrid and the dodecagrid, these tilings being classically denoted by $\{5,4\}$,
$\{7,3\}$ and $\{5,3,4\}$ respectively.
\par}

\section{Introduction}
\label{intro}
   In this paper, we look at the possibility to embed one-dimensional cellular automata,
$1D$- for short, into hyperbolic cellular automata in the pentagrid, the heptagrid or 
the dodecagrid which are denoted by $\{5,4\}$, $\{7,3\}$ and $\{5,3,4\}$ respectively. 
We consider $1D$-cellular automata which are deterministic and whose number of cells is 
infinite. 

   First, we shall prove a general theorem, and then we shall try to strengthen it
at the price of a restriction on the set of cellular automata which we wish to
embed in the case of the pentagrid.

   The first theorem says:

\begin{thm}\label{general}
There is a uniform algorithm to transform a deterministic $1D$-cellular automaton
with $n$~states into a deterministic cellular automaton in the pentagrid, the heptagrid or
the dodecagrid with, in each case, $n$$+$$1$~states. Moreover, the cellular automaton
obtained by the algorithm is rotation invariant.
\end{thm}

   Later on, as we consider deterministic cellular automata only, we drop this precision.
This theorem has a lot of corollaries, in particular we get this one, about
weak universality:

\begin{cor}\label{weakuniv3}
There is a weakly universal cellular automaton in the pentagrid, in the heptagrid
and in the dodecagrid which is weakly universal and which has three states exactly,
one state being the quiescent state. Moreover, the cellular automaton is rotation invariant.
\end{cor}

   We prove Theorem~\ref{general} and Corollary~\ref{weakuniv3} in Section~\ref{n+1}. 
In particular, we remind the notion of rotation invariance, especially for the
$3D$~case. In Section~\ref{n}, we strengthen the results, but this needs a restriction on 
the cellular automata under consideration in the case of the pentagrid. 

\section{Proof of Theorem~\ref{general} and its corollary}
\label{n+1}

   The idea of theorem~\ref{general} is very simple. Consider a one-dimensional
cellular automaton~$A$. The support of the cells of~$A$ is transported into a
structure of the hyperbolic grid which we consider as a {\bf line} of tiles. In each 
one of the three tilings which we shall consider, we define the line of tiles in 
a specific way. We examine these case, one after the other.

\subsection{In the pentagrid}
\label{line_penta_n+1}

In the case of the pentagrid, 
it is the set of cells such that one side of the cells is supported by the same
line of the hyperbolic plane which we assume to be a line of the tiling, call it the
{\bf guideline} of the implementation. It is a line supported by a side of a cell 
fixed once for all, see Figure~\ref{penta_n+1}. In this figure, the line is 
represented by the yellow cells along the guideline. Note that a yellow cell has
exactly two yellow neighbours. The cells are generated by the 
shift along the guideline which transforms one of the neighbours of the cell into
the cell itself. 

\vskip 10pt
\vtop{
\vspace{-10pt}
\setbox110=\hbox{\epsfig{file=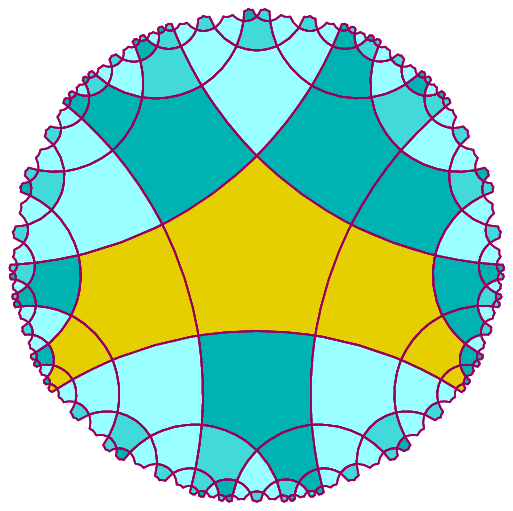,width=240pt}}
\ligne{\hfill
\PlacerEn {-125pt} {0pt} \box110
\hfill
}
\vspace{-5pt}
\begin{fig}
\label{penta_n+1}
\leurre
Implementation of a cellular automaton in the pentagrid.
The yellow cells represent the line of tiles used for the $1D$-CA.
The blue cells represent the cells which receive the new state.
\end{fig}
}
\vskip 10pt
In the figure, the yellow colour is assumed to represent the $n$~states of the 
original automaton. The blue cells represent the additional state which is different 
from the $n$~original ones.

   In the figure, there are three hues of blue which allow us to represent the tree
structure of the tiling. These different hues represent the same state.

   From the figure, it is plain that we have the following situation: yellow cells have 
exactly two yellow cells among their neighbours, the cell itself not being taken into
account. A blue cell has at most one yellow cell in its neighbourhood. Accordingly, this
difference is enough to define the implementation of the rules in the pentagrid.

Denote the format of a rule by $\eta_0\eta_1...\eta_5\eta_0^1$ where $\eta_0$ is the
current state of the cell, $\eta_i$ is the current state of neighbour~$i$ of the cell
and $\eta_0^1$ is the new state of the cell, obtained after the rule was applied.
We assume that the rules are {\bf rotation invariant}. This means that if $\pi$ is
a circular permutation on $\{1..5\}$ and if \hbox{$\eta_0\eta_1...\eta_5\eta_0^1$} is a rule
of the automaton, \hbox{$\eta_0\eta_{\pi(1)}...\eta_{\pi(5)}\eta_0^1$} is also a rule
of the automaton. As we assume the rules  to be invariant, the numbering has only 
to be fixed according to the orientation: we consider that it increases from~1 to~5 
as we clockwise turn around the tile. Which side is number~1 is not important. However, 
for the convenience of the reader, we fix it as follows.

   We may assume that the central cell has coordinate~0. We number its sides from~1 to~5,
1~being the number of the sides which is shared by a yellow cell on the left-hand side 
of the figure, the other sides being increasingly numbered while clockwise turning
around the tile. The left-hand part of the yellow line is the right-most branch of 
the tree which spans the corresponding quarter attached to the central cell, 
see~\cite{mmbook1,mmbook2} for explanations. In this cells, number~1 is given to the
side shared with the father and the others are also defined as for the central cells. In
particular, all the cells have their side~2 supported by the line defining the yellow
cells. On the right-hand side of the central cell, we have a branch defined by 
the middle son of the white nodes, starting from the root of the corresponding tree.
We take the same convention for the numbering of the sides, number~1 being given to the
side shared with the father. Then, we notice that all the cells of this part of the yellow
line have their side~5 on the line.

   Now, the rules for a blue cell are: \hbox{{\bf b}$\eta_1..\eta_5${\bf b}},
all $\eta_i$'s being~{\bf b} except possibly one of them.  The rules for a yellow
cell are: \hbox{$\eta_0\eta_1${\bf bb}$\eta_4${\bf b}$\eta_0^1$}, where
$\eta_1\eta_0\eta_4 \rightarrow \eta_0^1$ is the unique rule of~$A$ which can be associated
to the cell. 

\subsection{In the heptagrid}
\label{line_hepta_n+1}

   Figure~\ref{hepta_n+1} illustrates the implementation in the case of the heptagrid.

   This time, the guideline is not a line of the tiling as the lines which support a side
cut each second tile they meet. However, it is possible to define a guideline by taking
the mid-point lines: it was proved in~\cite{mmbook1} that the mid-points of two contiguous
sides of a heptagon define a line which cuts the other tiles at the mid-points of
two contiguous sides. It is not difficult to see that exactly two neighbours of a cell
crossed by the guideline are also crossed by this line, in the same way, through the 
mid-points of two consecutive sides and are in the same half-plane defined by the guideline.
By taking the shift along the guideline which transforms one of these neighbours into
the cell itself we can generate all the cells which belong to the expected line.

   In the setting of the heptagrid, the format of a rule is defined in terms which are 
very similar to those used for the pentagrid. The main difference is that here, the 
interval $[1..5]$ is replaced by $[1..7]$. The numbering of the sides is defined in the
same way as in the pentagrid. Indeed, as known from~\cite{mmbook1}, the pentagrid and
the heptagrid are spanned by the same tree. Again, fixing number~1 to the side shared with 
the father and a side, fixed once for all, in the case of the central cell, the cells which
are above the central cell have their sides~2 and~3 meeting the guideline and the cells which
are below have theirs side~6 and~7 meeting the guideline.

\vskip 10pt
\vtop{
\vspace{-10pt}
\setbox110=\hbox{\epsfig{file=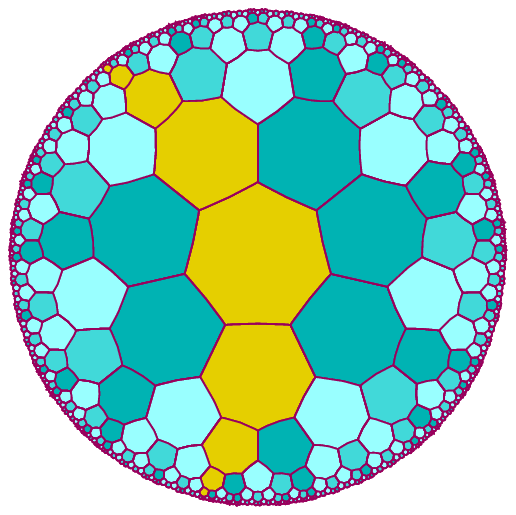,width=240pt}}
\ligne{\hfill
\PlacerEn {-125pt} {0pt} \box110
\hfill
}
\vspace{-5pt}
\begin{fig}
\label{hepta_n+1}
\leurre
Implementation of a cellular automaton in the heptagrid.
The yellow cells represent the line of tiles used for the $1D$-CA.
The blue cells represent the cells which receive the new state.
\end{fig}
}

Also, this time, the format of the rule is \hbox{$\eta_0\eta_1...\eta_7\eta_0^1$}.
Now, looking at Figure~\ref{hepta_n+1}, we can see that the rules for the blue
cells are of the form \hbox{{\bf b}$\eta_1...\eta_7${\bf b}} where, among
\hbox{$\eta_1...\eta_7$} two states exactly are {\bf b}. Moreover, these states are in
consecutive neighbours of the cell, taking into account the circular structure of the
neighbouring. This means that there is a circular permutation of the numbers such that
$\eta_1=\eta_2=\hbox{\bf b}$. Now, for a yellow cell, the format of the rules
is \hbox{$\eta_0\eta_1\hbox{\bf bbb}\eta_5\hbox{\bf bb}\eta_0^1$} for the central cell
together with the cells which are below, and
\hbox{$\eta_0\eta_1\hbox{\bf bb}\eta_4\hbox{\bf bbb}\eta_0^1$} for the cells above the
central cell. The central cell and those which are below apply the rule
\hbox{$\eta_1\eta_0\eta_5\rightarrow\eta_0^1$} of the automaton.
The cells which are above the central cell
apply the rule 
\hbox{$\eta_4\eta_0\eta_1\rightarrow\eta_0^1$}
of the automaton. 

   Accordingly, we proved Theorem~\ref{general} for what are the grid of the hyperbolic 
plane which we considered. It can easily be proved that the same result holds for
all the grids of the hyperbolic plane of the form $\{p,4\}$ and $\{p$+$2,3\}$, with
$p\geq 5$.

\subsection{In the dodecagrid}
\label{line_dodec_n+1}

   In the dodecagrid, we use the representation introduced in~\cite{mmarXiv3}.
We briefly remind it here for the convenience of the reader.

   In fact, we consider the projection of the dodecahedra on a plane which is defined 
by a fixed face of one of them: this will be the plane of reference~$\Pi_0$.
The trace of the tiling on~$\Pi_0$ is a copy of the pentagrid. So that, using a projection
of each dodecahedron which is in contact with~$\Pi_0$ and on the same half-space it defines
which we call the half-space {\bf above}~$\Pi_0$, we obtain a representation of the line
which is given by Figure~\ref{dodec_n+1}. Indeed, the projection of each dodecahedron on 
this face looks like a Schlegel diagram, see~\cite{mmarXiv3,mmbook1} for more details
on this tool dating from the 19$^{\rm th}$ century. 
\vskip 10pt
\vtop{
\vspace{-10pt}
\setbox110=\hbox{\epsfig{file=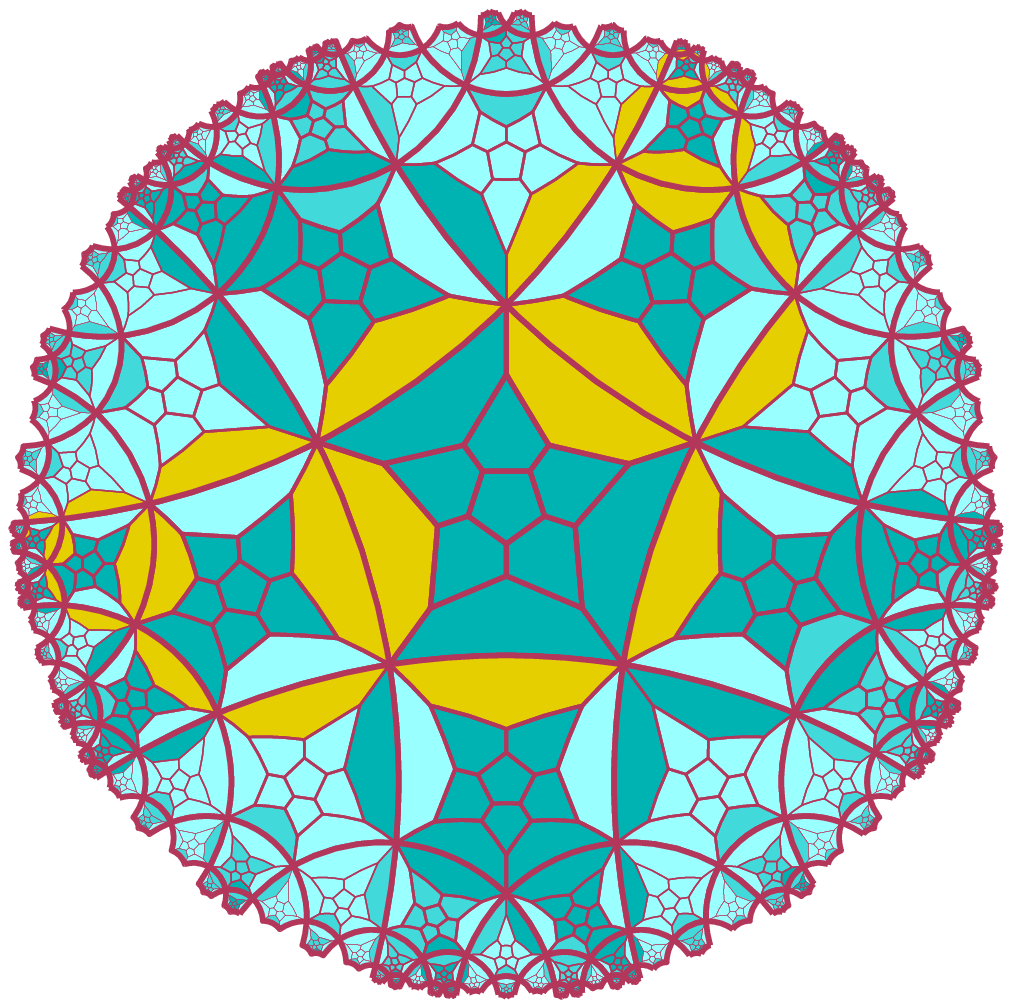,width=240pt}}
\ligne{\hfill
\PlacerEn {-125pt} {0pt} \box110
\hfill
}
\vspace{-5pt}
\begin{fig}
\label{dodec_n+1}
\leurre
Implementation of a cellular automaton in the dodecagrid.
The yellow cells represent the line of tiles used for the one-dimensional CA.
The blue cells represent the cells which receive the new state.
\end{fig}
}

   Accordingly, the guideline is simply a line of the pentagrid which lies on~$\Pi_0$.
On the figure, we can see that the line which implements the one-dimensional cellular
automaton is represented by the yellow cells, the other cells which receive the new state
being blue. This line of yellow cells will be also called the {\bf yellow line}. As 
in Figures~\ref{penta_n+1} and~\ref{hepta_n+1}, the different hues of blue are used in order
to show the spanning trees of the pentagrid, dispatched around the
central cell.

   To define the rules of a cellular automaton, we also introduce a numbering of the
faces of a dodecahedron which will allow us to number the neighbours. This numbering
is given by Figure~\ref{numdodec}.

\vskip 10pt
\vtop{
\vspace{-10pt}
\setbox110=\hbox{\epsfig{file=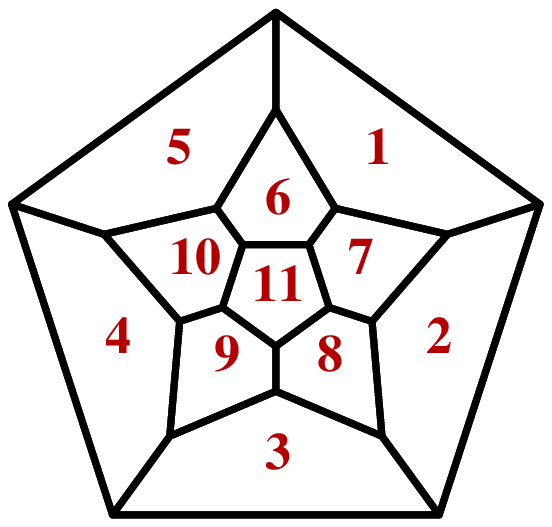,width=240pt}}
\ligne{\hfill
\PlacerEn {-125pt} {0pt} \box110
\hfill
}
\vspace{-5pt}
\begin{fig}
\label{numdodec}
\leurre
The numbering of the faces of dodecahedron. Face~$0$ is delimited by the biggest pentagon
of the figure.
\end{fig}
}

   Accordingly, the format of a rule is of the form
\hbox{$\eta_0\eta_1...\eta_{12}\eta_0^1$}. Now, as the rules are assumed to be rotation
invariant, which face receives number~1 is not important. However, for the convenience
of the reader, we shall adopt the following convention. For all the yellow cells, 
we consider that the face which is on~$\Pi_0$ is face~0. Accordingly, the numbers 
should appear in Figure~\ref{dodec_n+1} as they appear
in Figure~\ref{numdodec}. Moreover, we consider that the other face of the cell which
is in contact with the guideline is face~5.

   Now, we take this occasion to remember how we can see whether two numberings
of the dodecahedron are obtained from one another by a positive displacement
in the space.

\subsubsection{Rotation invariance}
\label{rotation_invariance}

   The question is the following: how does a motion which leaves the dodecahedron
globally invariant affect the numbering of its faces, an initial numbering being fixed
as in Figure~\ref{numdodec}?

   In fact, it is enough to consider products of rotations as we do not consider
reflections in planes. The simplest way to deal with this problem is the following.
Consider a motion which preserves the orientation, we shall say a {\bf positive} motion. 
As it leaves the dodecahedron globally invariant, it transforms the face into another 
one. Accordingly, fix face~0. Then its image can be any face of the dodecahedron, 
face~0 included. Let $f_0$ be the image of face~0. Next, fix a second face which shares 
an edge with face~0, for instance face~1. Then its image $f_1$ is a face which shares 
an edge with~$f_0$. It can be any face sharing a face with~$f_0$. 
Indeed, let $f_2$ be another face sharing an edge with~$f_1$. Then, composing 
the considered positive motion with a rotation around $f_0$ transforming $f_1$ into~$f_2$, 
we get a positive motion which transforms $(0,1)$ into $(f_0,f_2)$. This proves that we 
get all the considered positive motion leaving the dodecahedron globally invariant,
by first fixing the image of face~0, say $f_0$ and then by taking any face~$f_1$
sharing an edge with~$f_1$. Note that once $f_0$ and~$f_1$ are fixed, the images of
the other faces are fixed, thanks to the preservation of the orientation.
Accordingly, there are 60 of these positive motions and the argument of the proof
shows that they are all products of rotations leaving the dodecahedron globally
invariant. 

\setbox110=\hbox{\epsfig{file=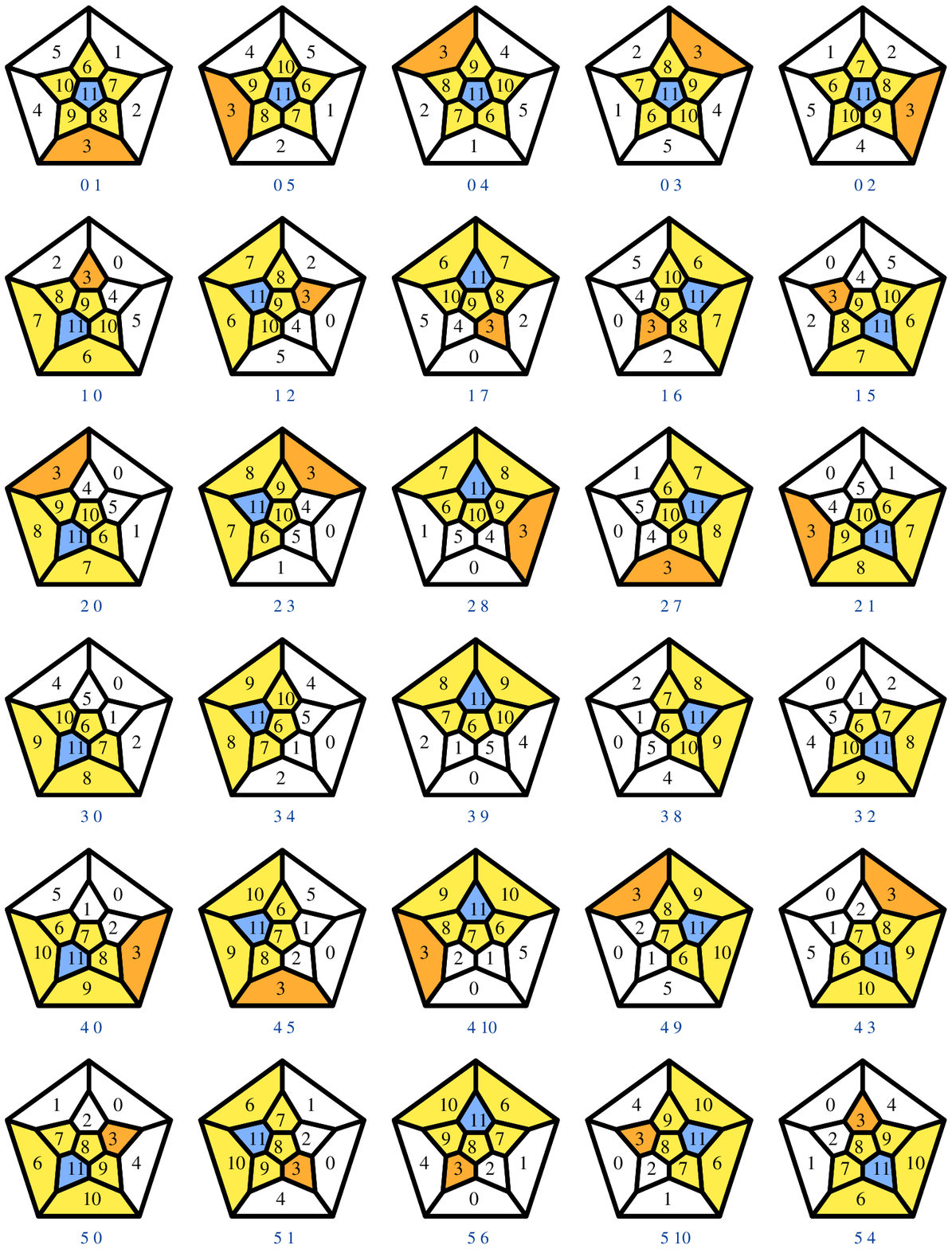,width=180pt}}
\setbox112=\hbox{\epsfig{file=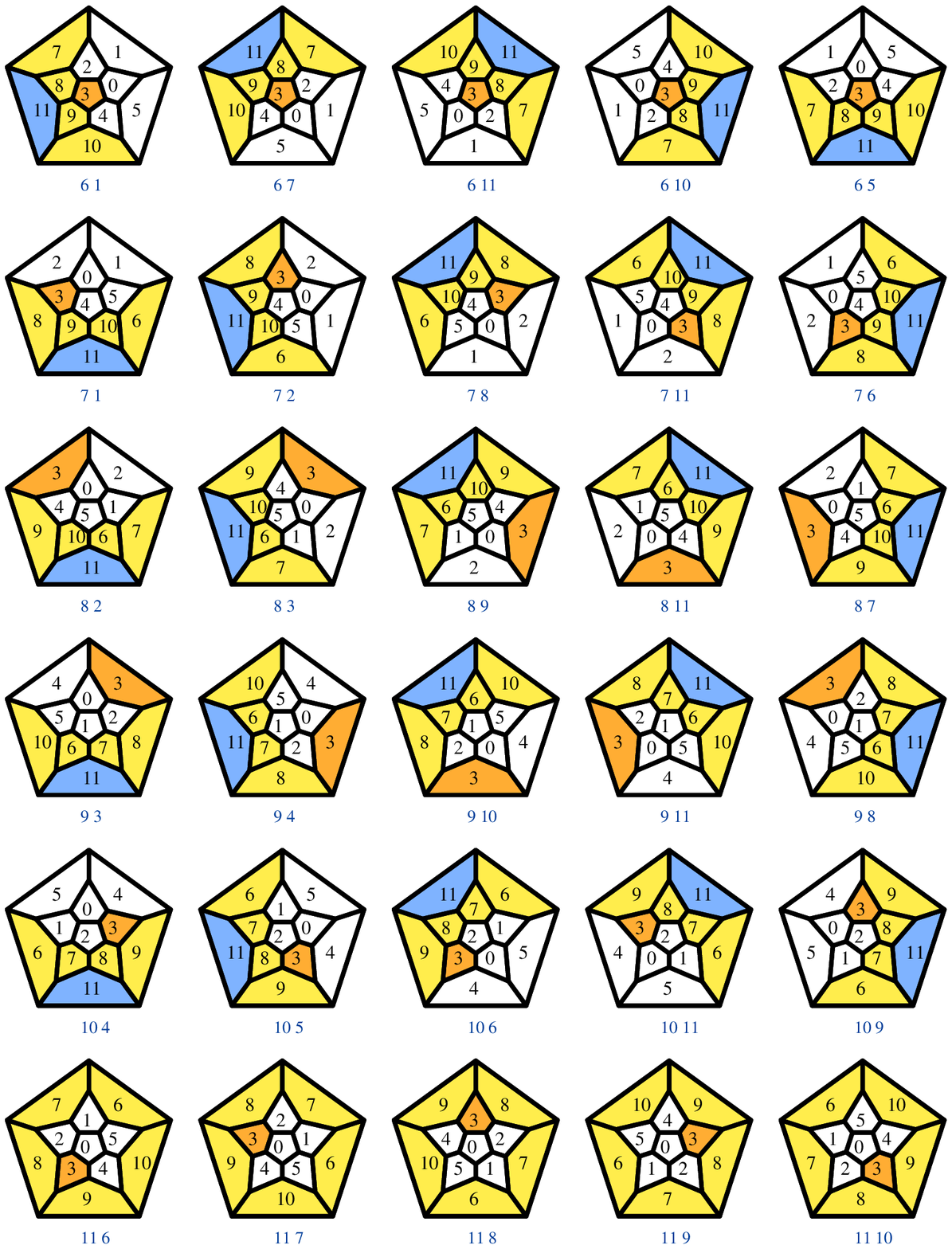,width=180pt}}
\vtop{
\vspace{-15pt}
\ligne{\hfill
\PlacerEn {-345pt} {0pt} \box110
\PlacerEn {-165pt} {0pt} \box112
}
\vspace{-20pt}
\begin{fig}\label{rotations}
\leurre
The map of the positive motions leaving the dodecahedron globally invariant.
\end{fig}
}
\vskip 7pt
   Figure~\ref{rotations} gives an illustrative classification of all these rotations.
The upper left picture represents the image of a Schlegel diagram of a dodecahedron
with the notation introduced in Figure~\ref{numdodec}. Each image represents 
a positive motion. Its characterization is given by the couple of numbers under the
image: it has the form~$f_0\ f_1$, where $f_0$ is the image of face~0 and $f_1$ is 
the image of face~1. The figure represents two sub-tables, each one containing
30~images. Each row represents the possible images of~$f_1$, $f_0$ being fixed.
The image of face~0 is the back of the dodecahedron. The image of face~1
is the place of face~1{} in Figure~\ref{numdodec}. As an example, $f_0=0$ 
for the first row of the left-hand side sub-table, and in the first row, the first
image gives $f_1=1$, so that it represents the identity. The other images
of the row represent the rotations around face~0.

\def\rangee #1 #2 #3 #4 #5 #6 {%
\ligne{\hbox to 30pt{\hfill#1\hfill}
       \hbox to 25pt{\hfill#2\hskip 7pt}
       \hbox to 25pt{\hfill#3\hskip 7pt}
       \hbox to 25pt{\hfill#4\hskip 7pt}
       \hbox to 25pt{\hfill#5\hskip 7pt}
       \hbox to 25pt{\hfill#6\hskip 7pt}
      }
\vskip 4pt
}
\vtop{
\begin{tab}\label{couronnes}
\leurre
The faces around a given face.
\end{tab}
\vspace{-12pt}
\ligne{\hfill
\vtop{\leftskip 0pt\parindent 0pt\hsize=190pt
\grostrait
\ligne{\hfill
       \vtop{\offinterlineskip\leftskip 0pt\parindent 0pt\hsize=168.3333pt
\rangee   { } 1  2   3   4   5
}\hfill}
\demitrait
\ligne{\hfill
       \vtop{\offinterlineskip\leftskip 0pt\parindent 0pt\hsize=168.3333pt
\rangee   0   1  5   4   3   2
\rangee   1   0  2   7   6   5
\rangee   2   0  3   8   7   1
\rangee   3   0  4   9   8   2
\rangee   4   0  5  10   9   3
\rangee   5   0  1   6  10   4
\rangee   6   1  7  11  10   5
\rangee   7   1  2   8  11   6
\rangee   8   2  3   9  11   7
\rangee   9   3  4  10  11   8
\rangee  10   4  5   6  11   9
\rangee  11   6  7   8   9  10
}
\hfill}
\demitrait
}
\hfill}
}
\vskip 10pt
The construction of Figure~\ref{rotations} was performed by an algorithm
using Table~\ref{couronnes}. For each face of the dodecahedron, the table gives
the faces which surround it in the Schlegel diagram, taking the clockwise order
when looking at the face from outside the dodecahedron, this order coinciding with
increasing indices in each row. This coincides with the 
usual clockwise order for all faces as in Figure~\ref{numdodec}, except for face~0
for which the order is counter-clockwise when looking above the plane of the projected
image. The principle of the drawings consists in placing~$f_0$ onto face~0 
and~$f_1$ onto face~1. The new numbers of the faces are computed by the algorithm
as follows. Being given the new numbers~$f_0$ and~$f_1$ of two contiguous 
faces~$\varphi_0$ and~$\varphi_1$ in the Schlegel diagram, the algorithm computes 
the position of $\varphi_1$ as a neighbour of~$\varphi_0$ in the table. This
allows to place~$f_1$ on the right face. Then, the algorithm computes the new
numbers of the faces which are around~$\varphi_1$ in the table: it is enough to take
the position of~$\varphi_0$ as a neighbour of~$\varphi_1$ and then to turn around
the neighbours of~$f_1$, looking at the new numbers in the row~$f_1$ of the table,
starting from the position of~$f_0$. This gives the new numbers of the faces
which surround face~1. It is easy to see that we have all faces
of the dodecahedron by turning around face~1, then around face~5, then around face~7 
and at last around face~8. As in these steps, each round of faces starts from a face 
whose new number is already computed, the algorithm is able to compute the new numbers 
for the current round of faces, using Table~\ref{couronnes} to find the
new numbers. Let us call this algorithm the {\bf rotation 
algorithm}.

   Thanks to the rotation algorithm, it is easy to compute the {\bf rotated forms}
of a rule of the cellular automaton.

\subsubsection{The rules in the dodecagrid}

   Let $\eta^0\eta_0...\eta_{11}\eta^1$ be a rule of the automaton: $\eta^0$~is
the current state of the cell, $\eta_i$ is that of the neighbour which the cell can
see through its face~$i$, $\eta^1$ is the new state of the cell.
Remember that the current state of a cell is its sate at time~$t$ and that its new state is
its state at time~$t$+1, after the rule was applied at time~$t$. We shall decide that
a neighbour is numbered by the number of the face through which it is seen by the cell.
By assumption, this numbering is a rotated image of the numbering defined by 
Figure~\ref{numdodec}. Later we shall call
$\eta^0\eta_0...\eta_{11}$ the {\bf context} of the rule.
Let $\mu$ be a positive motion leaving the dodecahedron globally invariant. 
The {\bf rotated form} of the rule defined by~$\mu$ is 
$\eta^0\eta_{\mu(0)}...\eta_{\mu(11)}\eta^1$ and, similarly,
$\eta^0\eta_{\mu(0)}...\eta_{\mu(11)}$ is the {\bf rotated form} by~$\mu$
of the context of the initial rule.
We say that the cellular automaton is {\bf rotation invariant} if and only
two rules having contexts which are rotated forms of each other always produce the same
new state.

   Now, thanks to our study, we have a syntactic criterion to check this property.
We fix an order of the states. Then, for each rule, we compute its {\bf minimal form}. 
This form is obtained as follows. We compute all rotated forms of the rule and,
looking at the obtained contexts as words, we take their minimum in the lexicographic 
order. The minimal form of a rule is obtained by appending its new state to this minimum. 
Now it is easy to see that:

\begin{lem}~{\rm (see~\cite{mmarXiv3})}
A cellular automaton on the dodecagrid is rotation invariant if and only if
for any pair of rules, if their minimal forms have the same context, they 
have the same new state too.
\end{lem}

   Now, checking this property can easily be performed thanks to the rotation algorithm.

   As we already indicated, we decided that face~0 of the cells belonging to the line
of the implementation are on~$\Pi_0$ and that the other face which has a side
on the guideline is face~5. As a consequence, a yellow cell is in contact with two yellow
neighbours by its faces~1 and~4. We decide that the face~1 of a cell is the same as the
face~4 of the next yellow neighbour and, accordingly, its face~4 is the same as the face~1 of
the other yellow neighbour. This allows to define two directions on the yellow line.
The direction from left to right on the one-dimensional cellular automaton is, by convention,the direction from face~1 to face~4 of the same cell.

   For the proof of Theorem~\ref{general} in the case of the dodecagrid, the rules 
for a blue cell have the form \hbox{${\bf b}\eta_1...\eta_{12}\hbox{\bf b}$} 
with all states in \hbox{$\eta_1...\eta_{12}$} being {\bf b} except, possibly, one of them.
From the just defined convention on the numbering of the faces of the yellow cells,
the rules for a yellow cell are of the form
\hbox{$\eta^0\hbox{\bf b}\eta_1\hbox{\bf bb}\eta_4\hbox{\bf bbbbbbb}\eta^1$},
where \hbox{$\eta_1\eta^0\eta_4\rightarrow\eta^1$} is the rule of the one-dimensional
cellular automaton.

   Now, as the blue cells have at most one yellow neighbour and as the yellow cells
have two yellow neighbours exactly, the difference between the rules is clearly
recognizable. 

   This completes the proof of Theorem~\ref{general}.\cqfd

   Now, the proof of Corollary~\ref{weakuniv3} is very easy: it is enough to apply
the theorem to the elementary cellular automaton defined by rule~110 which is now known
to be weakly universal, see~\cite{cook, wolfram}.

\section{Refinement of Theorem~\ref{general}}
\label{n}

   Now, we shall prove that, under particular hypotheses in the case of the
pentagrid and no restriction in the case of the heptagrid and of the dodecagrid, 
a $1D$ cellular automaton with $n$~states can be simulated by 
hyperbolic cellular automaton with $n$~states too. 

   In order to formulate this hypothesis, consider a one-dimensional deterministic
cellular automaton~$A$. Say that a state~$s$ of~$A$ is {\bf fixed} in the context $x,y$ 
in this order, if the rule $xsy\rightarrow s$ belongs to the table of transitions of~$A$. 
As an example, a {\bf quiescent} state for~$A$, usually denoted by~0, is fixed in the 
context $0,0$. Now, we say that $A$~is a {\bf fixable} cellular automaton if it has 
a quiescent state~0 which is also fixed in the context $1,0$, and another state, 
denoted by~1, such that 1~is fixed in the context $0,0$.

   We can now formulate the following results:

\begin{thm}\label{reffinment1}
There is an algorithm which transforms any fixable $1D$ cellular automaton~$A$
with $n$~states into a rotation invariant cellular automaton~$B$ in the pentagrid 
with $n$~states too, such that $B$~simulates $A$ on a line of the pentagrid. 
\end{thm}

\begin{thm}\label{reffinment2}
There is an algorithm which transforms any deterministic $1D$ cellular 
automaton~$A$ with $n$~states into a rotation invariant deterministic cellular 
automaton~$B$ in the  heptagrid, the dodecagrid respectively, with $n$~states too,
such that $B$~simulates $A$ on a line of the heptagrid, the dodecagrid respectively. 
\end{thm}

   First, we prove Theorem~\ref{reffinment1}.

   To this purpose, we consider Figure~\ref{penta_n}. In this figure, the yellow colour
is still used to represent any state of the automaton~$A$. Now, the green colour 
represents the quiescent state~$0$, and the red one represents the state~1 which is
fixed in the context $0,0$. We also assume that $0$~is fixed in the context $1,0$. 

   We shall consider all the neighbours of the central cell. Its red neighbour will 
be numbered by~1, and the others from~2 to~5, increasing as we clockwise turn around the
cell. We also consider the cells which just has one vertex in common with the central cell.
All the other cells are in quiescent state or they belong to the yellow line or are 
neighbouring a cell belonging to this line. In this latter case, such a cell is obtained 
from one of those we consider around the central cell by a shift along the guideline. 

   Define $B$ with $n$~states represented by different letters from those used from~$A$.
We fix a bijection between the states of~$A$ and those of~$B$ in which {\bf B} is
associated to the state~1 of~$A$ and {\bf W} is associated to the state~0 of~$A$.

\newdimen\largee\largee=18pt
\newdimen\largea\largea=25pt
\def\larangee #1 #2 #3 #4 #5 #6 #7 {%
\ligne{\hbox to \largea{\hskip 5pt#1\hfill}
       \hbox to \largee{\hfill#2\hfill}
       \hbox to \largee{\hfill#3\hfill}
       \hbox to \largee{\hfill#4\hfill}
       \hbox to \largee{\hfill#5\hfill}
       \hbox to \largee{\hfill#6\hfill}
       \hbox to \largee{\hfill#7\hfill}
}
}

   Consider the configuration around the central cell. If we write the states of the cell
and then those of its neighbours according to the order of their numbers, we get the 
following word:
\hbox{\bf $Y$BW$Z$W$X$}, where $X,Y,Z$ are taken among the states of~$B$.
Now, if $Z$~=~{\bf B}, we can start from this neighbour in state~{\bf B} which has 
number~3, and we get the word \hbox{\bf $Y$BW$X$BW} in which we see {\bf B} in position~4. 
If $X$~=~{\bf B}, then we get the word \hbox{\bf $Y$BBW$Z$W} in which we see {\bf B} in
position~2. In both case, the configuration around the cell is different from the one we 
obtain by starting from position~1. We can synthesise this information as follows:

\vskip 5pt
\ligne{\hfill
\vtop{\hsize=148.56pt
\grostrait
\larangee {} {} 1 2 3 4 5
\vskip-3pt
\demitrait
\larangee 0 {$Y$} {\bf B} {\bf W} {$Z$} {\bf W} {$X$}
\larangee {} {} {\bf B} {\bf W} {$X$} {\bf B} {\bf W} 
\larangee {} {} {\bf B} {\bf B} {\bf W} {$Z$} {\bf W}
\vskip-6pt
\demitrait
}
\hfill
}
\vskip 10pt
The first line corresponds to the configuration which triggers the application of
the rule of~$A$ corresponding to \hbox{$XYZ \rightarrow X'$}. Clearly, as already noticed
with the positions of the fixed {\bf B} and~{\bf W}, the other lines do
not correspond to the application of a rule of~$A$.

  We shall do this for all the neighbours of the central cell, and in Table~\ref{releve5},
we can see all the possible configurations for the neighbours of the central cell.

\vtop{
\begin{tab}\label{releve5}
Table of the configurations around the central cell in the pentagrid for the automaton~$B$.
\end{tab}
\vspace{-12pt}
\ligne{\hfill
\vtop{\hsize=148.56pt
\grostrait
\vskip 3pt
\larangee {} {} 1 2 3 4 5
\vskip -3pt
\demitrait
\larangee 0 {$Y$} {\bf B} {\bf W} {$Z$} {\bf W} {$X$}
\larangee {} {} {\bf B} {\bf W} {$X$} {\bf B} {\bf W} 
\larangee {} {} {\bf B} {\bf B} {\bf W} {$Z$} {\bf W}
\vskip-6pt
\demitrait
\larangee {$1_1$} {\bf B} {$Y$} {\bf W} {\bf W} {\bf W} {\bf W}
\vskip-6pt
\demitrait
\larangee {$2_1$} {\bf W} {\bf B} {\bf W} {\bf W} {\bf W} {\bf W}
\vskip-6pt
\demitrait
\larangee {$1_2$} {\bf W} {$Y$} {\bf W} {\bf W} {\bf W} {\bf B}
\larangee {} {} {\bf B} {\bf W} {\bf W} {\bf W} {\bf B}
\larangee {} {} {\bf B} {\bf B} {\bf W} {\bf W} {\bf B}
\larangee {} {} {\bf B} {\bf W} {\bf W} {\bf W} {\bf W}
\vskip-6pt
\demitrait
\larangee {$2_2$} {\bf B} {\bf W} {\bf W} {\bf W} {\bf W} {$Z$}
\larangee {} {} {\bf B} {\bf W} {\bf W} {\bf W} {\bf W}
\vskip-6pt
\demitrait
\larangee {$1_3$} {$Z$} {$Y$} {\bf B} {\bf W} {$T$} {\bf W}
\larangee {} {} {\bf B} {\bf B} {\bf W} {$T$} {\bf W} 
\larangee {} {} {\bf B} {\bf W} {$T$} {\bf W} {$Y$}
\larangee {} {} {\bf B} {\bf W} {$Y$} {\bf B} {\bf W}
\vskip-6pt
\demitrait
}
\hfill
\vtop{\hsize=148.56pt
\grostrait
\vskip 3pt
\larangee {} {} 1 2 3 4 5
\vskip -3pt
\demitrait
\larangee {$2_3$} {\bf W} {$Z$} {\bf W} {\bf W} {\bf W} {\bf W} 
\vskip-6pt
\demitrait
\larangee {$1_4$} {\bf W} {$Y$} {\bf W} {\bf W} {\bf W} {\bf W}
\vskip-6pt
\demitrait
\larangee {$2_4$} {\bf W} {\bf W} {\bf W} {\bf W} {\bf W} {$X$}
\larangee {} {} {\bf B} {\bf W} {\bf W} {\bf W} {\bf W}
\vskip-6pt
\demitrait
\larangee {$1_5$} {$X$} {$Y$} {\bf W} {$U$} {\bf B} {\bf W}
\larangee {} {} {\bf B} {\bf W} {$U$} {\bf B} {\bf W}
\larangee {} {} {\bf B} {\bf B} {\bf W} {$Y$} {\bf W}
\larangee {} {} {\bf B} {\bf W} {$Y$} {\bf W} {$U$}
\vskip-6pt
\demitrait
\larangee {$2_5$} {\bf W} {$X$} {\bf W} {\bf W} {\bf W} {\bf B}
\larangee {} {} {\bf B} {\bf W} {\bf W} {\bf W} {\bf B}
\larangee {} {} {\bf B} {\bf B} {\bf W} {\bf W} {\bf W}
\larangee {} {} {\bf B} {\bf W} {\bf W} {\bf W} {\bf W}
\vskip-6pt
\demitrait
}
\hfill}
}
\vskip 10pt
   In Table~\ref{releve5}, we indicate the coordinate of the cell which we represent
together with its state. Then, if there are states as $U$, $X$, $Y$, $Z$, $T$, we 
also represent the case when one of this variable takes the value~{\bf B} and we 
represent the configuration around the cell when this~{\bf B} is put onto position~1.

\vskip 10pt
\vtop{
\vspace{-10pt}
\setbox110=\hbox{\epsfig{file=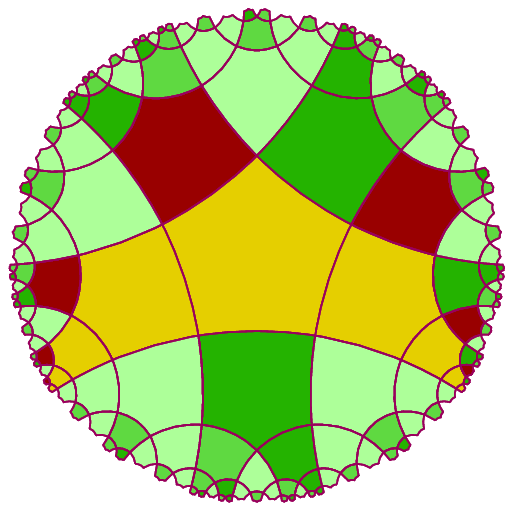,width=240pt}}
\ligne{\hfill
\PlacerEn {-125pt} {0pt} \box110
\hfill
}
\vspace{-5pt}
\begin{fig}
\label{penta_n}
\leurre
Implementation of a $1D$ cellular automaton in the pentagrid.
The yellow cells represent the line of tiles used for the $1D$-CA.
The green cells represent the cells which receive a particular state among the states
of the $1D$-CA.
\end{fig}
}
\vskip 10pt
   Figure~\ref{penta_n} allows us to check the correctness of Table~\ref{releve5}.
Now, looking at the table, we can see that there are several cases. In the first one,
the table displays one configuration only. This is the case for cell~$2_1$ for instance.
Its state is~{\bf W} and the configuration around it is \hbox{\bf BWWWW}, with {\bf B}
in position~1. This configuration is compatible with the application of a rule of~$A$.
From the first row of the table associated to the central cell, we can see that the
states corresponding to a rule of~$A$ are those which lie in the guideline, namely
the states in positions~3 and~5 in this order: position~5 as the left-hand side neighbour
and position~3 as the right-hand side neighbour. We shall denote this as the
configuration $5\__o\_3$. As cell $2_1$ does not belong to the guideline, its state must
be unchanged. This requires the rule \hbox{\bf WWW $\rightarrow$ W}, which exists as
0~is a quiescent state for~$A$. We have a similar situation for cells~$2_3$ and~$1_4$ 
when state~$Z$ is~{\bf B} for cell~$2_3$ and when state~$Y$ is~{\bf B} for cell~$1_4$.
We remain with one cell with a single configuration: cell~$1_1$ in state~{\bf B} with 
the configuration \hbox{\bf $Y$WWWW}. When $Y$~is {\bf B}, we have that the rules of~$A$
can be applied and the configuration $5\__o\_3$ is now: \hbox{\bf WBW}, and the state
must remain unchanged, requiring the rule \hbox{\bf WBW $\rightarrow$ B}. Now,
this rule exists in the table of transition of~$A$ as $A$~is fixable: state~1 is fixed
for the configuration $0,0$. 

   Now, let us consider the cells~$2_2$ and~$2_4$. In these cases, there is a variable
state in position~5, the others being~{\bf W}. If the variable takes the value~{\bf B},
then, taking this {\bf B}~in position~1, we get a situation corresponding to the cases
of cell~$2_1$ or~$1_1$.

   Next, consider the cells~$1_2$ and~$2_5$, where we have the same situation, exactly. 
In the configuration of the first line, we get the next one if $Y$ is~{\bf B} in the
case of cell~$1_2$, when $X$ is~{\bf B} in the case of cell~$2_5$. This configuration
allows the application of a rule of~$A$. The configuration $5\__o\_3$ being
\hbox{\bf BWW} and the state of the cell having to remain unchanged, this requires
the rule \hbox{\bf BWW $\rightarrow$ W}. This corresponds to the rule 
\hbox{100 $\rightarrow$ 0} which belongs to the transitions of~$A$ as 0~is stable
for the configuration~$1,0$. Now, if we take the state~{\bf B} of the initial configuration
in cells~$1_2$ and~$2_5$ in position~1, we get a configuration which is compatible with
the application of a rule of~$A$ when $Y$ or~$X$ is~{\bf W}, otherwise we get {\bf B} in
position~2 also which bars the application of a rule of~$A$. Now, in this case, we have
the situation which was already analysed in the case of cell~$2_1$ for instance.  

   At last, we remain with the cells~$1_3$ and~$1_5$. Note that if we put the state~$U$
in position~1 in the case of cell~$1_5$, we have exactly the same configuration as the
first line of the table for cell~$1_3$. Accordingly, it is enough to look at cell~$1_3$.
Now, the first configuration does not allow an application of a rule of~$A$ as we have
{\bf B} in position~2, which is illustrated by the second line of this entry of the table.
Now, if we put this {\bf B} in position~1, this is illustrated by the third line,
we get a configuration which requires a rule of the automaton~$A$. Considering 
the configuration~$5\__o\_3$ and the state of the cell being~$Z$, we
need the rule \hbox{$YZT \rightarrow Z'$} which is a rule of~$A$, by assumption, as
cell~$1_3$ belongs to the yellow line. It remains to see at what happens when $T$ 
is~{\bf B} if this {\bf B}~is put in position~1. As shown by the forth line, the 
initial~{\bf B} then occurs in position~4 which bars an application of a rule of
automaton~$A$. 

   Now, from this study, we can see that for each cell around the central cell and for
this cell also, there is at most one configuration around the cell which is
compatible with the application of a rule of~$A$. In fact, the single case which we 
did not study is a cell in state~{\bf W} which does not belong to the yellow line and 
whose neighbours are also in state~{\bf W}. Such a configuration is not compatible
with the application of a rule of~$A$ and, of course, we decide that in this case, the
state of the cell is unchanged. From our study, we also have seen that when a rule
of~$A$ can be applied to a cell which does not belong to the yellow line, then
the rule never changes the state of the cell, thanks to the hypothesis that $A$~is fixable.

   And so, $B$ works as follows: if around the cell there is a configuration which is
compatible with the application of a rule of~$A$, the state is changed according to this
rule of~$A$, otherwise the state is not changed.

   This completes the proof of Theorem~\ref{reffinment1}.

\newdimen\largeb\largeb=16pt
\newdimen\largec\largec=13.5pt
\def\larangea #1 #2 #3 #4 #5 {%
\hbox to \largeb{\hskip 2pt#1\hfill}
       \hbox to \largec{\hfill#2\hfill}
       \hbox to \largec{\hfill#3\hfill}
       \hbox to \largec{\hfill#4\hfill}
       \hbox to \largec{\hfill#5\hfill}
}
\def\larangeb #1 #2 #3 #4 {%
       \hbox to \largec{\hfill#1\hfill}
       \hbox to \largec{\hfill#2\hfill}
       \hbox to \largec{\hfill#3\hfill}
       \hbox to \largec{\hfill#4\hfill}
}

\vskip 10pt
   Now, let us turn to the case of the heptagrid. In this case, the situation is in some 
sense easier as it requires no special hypothesis on the deterministic $1D$ 
cellular automaton. Indeed, the fact is that due to the number of neighbours, there is
a way to differentiate the cells belonging to the yellow line from those which do not.
As mentioned in Subsection~\ref{line_hepta_n+1}, the yellow line is now implemented
along a mid-point line of the heptagrid which is fixed, once for all. As in the case
of the pentagrid, the yellow colour represents any state of automaton~$A$. Now,
we assume that~$A$ has at least two states, 0 and~1. In Figure~\ref{hepta_n}, these
state are represented in green and in red respectively. As in the case of the pentagrid, 
we use different hues of green in order to make visible the tree structure which spans
the tiling. Now, it is easy to see that the configurations allowing the application
of a rule of~$A$ are reached only in the case of cells of the yellow line and that for 
these cells, among the rotated contexts, exactly one is compatible with the application
of a rule of~$A$. This can be checked on the figure and we report this examination
in Table~\ref{releve7}. 
    
\vtop{
\begin{tab}\label{releve7}
Table of the configurations around the central cell in the pentagrid for the automaton~$B$.
\end{tab}
\vspace{-12pt}\rmix\mathix
\ligne{\hfill
\vtop{\hsize=163.36pt
\grostrait
\vskip 3pt
\ligne{\larangea {} {} 1 2 3 
\larangeb 4 5 6 7 }
\vskip -3pt
\demitrait
\ligne{
\larangea 0 {$Y$} {$X$} {\bfix B} {\bfix W} 
\larangeb {\bfix B} {$Z$} {\bfix W} {\bfix W} }  
\ligne{
\larangea {} {} {\bfix B} {\bfix B} {\bfix W} 
\larangeb {\bfix B} {$Z$} {\bfix W} {\bfix W} }  
\ligne{
\larangea {} {} {\bfix B} {\bfix W} {\bfix B} 
\larangeb {$Z$} {\bfix W} {\bfix W} {$X$} } 
\ligne{
\larangea {} {} {\bfix B} {$Z$} {\bfix W} 
\larangeb {\bfix W} {$X$} {\bfix B} {\bfix W } } 
\ligne{
\larangea {} {} {\bfix B} {\bfix W} {\bfix W} 
\larangeb {$X$} {\bfix B} {\bfix W } {\bfix B} } 
\vskip-6pt
\demitrait
\ligne{
\larangea {$1_1$} {$X$} {$Y$} {\bfix W} {\bfix W} 
\larangeb {$U$} {\bfix B} {\bfix W} {\bfix B} }  
\ligne{
\larangea {} {} {\bfix B} {\bfix W} {\bfix W} 
\larangeb {$U$} {\bfix B} {\bfix W} {\bfix B} }  
\ligne{
\larangea {} {} {\bfix B} {\bfix B} {\bfix W} 
\larangeb {\bfix B} {$Y$} {\bfix W} {\bfix W} }  
\ligne{
\larangea {} {} {\bfix B} {\bfix W} {\bfix B} 
\larangeb {$Y$} {\bfix W} {\bfix W} {$U$} }  
\ligne{
\larangea {} {} {\bfix B} {$Y$} {\bfix W} 
\larangeb {\bfix W} {$U$} {\bfix B} {\bfix W} }  
\vskip-6pt
\demitrait
\ligne{
\larangea {$1_2$} {\bfix B} {$Y$} {$X$} {\bfix W} 
\larangeb {\bfix W} {\bfix W} {\bfix W} {\bfix W} }  
\ligne{
\larangea {} {} {\bfix B} {$X$} {\bfix W} 
\larangeb {\bfix W} {\bfix W} {\bfix W} {\bfix W} }  
\ligne{
\larangea {} {} {\bfix B} {\bfix W} {\bfix W} 
\larangeb {\bfix W} {\bfix W} {\bfix W} {$Y$} }  
\vskip-6pt
\demitrait
\ligne{
\larangea {$1_3$} {\bfix W} {$Y$} {\bfix B} {\bfix W} 
\larangeb {\bfix W} {\bfix W} {\bfix W} {\bfix B} }  
\ligne{
\larangea {} {} {\bfix B} {\bfix B} {\bfix W} 
\larangeb {\bfix W} {\bfix W} {\bfix W} {\bfix B} }  
\ligne{
\larangea {} {} {\bfix B} {\bfix W} {\bfix W} 
\larangeb {\bfix W} {\bfix W} {\bfix B} {$Y$} }  
\vskip-6pt
\demitrait
}
\hfill
\vtop{\hsize=163.36pt
\grostrait
\vskip 3pt
\ligne{\larangea {} {} 1 2 3 
\larangeb 4 5 6 7 }
\vskip -3pt
\demitrait
\ligne{
\larangea {$1_4$} {\bfix B} {$Y$} {\bfix W} {\bfix W} 
\larangeb {\bfix W} {\bfix W} {\bfix W} {$Z$} }  
\ligne{
\larangea {} {} {\bfix B} {\bfix W} {\bfix W} 
\larangeb {\bfix W} {\bfix W} {\bfix W} {$Z$} }  
\ligne{
\larangea {} {} {\bfix B} {$Y$} {\bfix W} 
\larangeb {\bfix W} {\bfix W} {\bfix W} {\bfix W} }  
\vskip-6pt
\demitrait
\ligne{
\larangea {$1_5$} {$Z$} {$Y$} {\bfix B} {\bfix W} 
\larangeb {\bfix B} {$T$} {\bfix W} {\bfix W} }  
\ligne{
\larangea {} {} {\bfix B} {\bfix B} {\bfix W} 
\larangeb {\bfix B} {$T$} {\bfix W} {\bfix W} }  
\ligne{
\larangea {} {} {\bfix B} {\bfix W} {\bfix B} 
\larangeb {$T$} {\bfix W} {\bfix W} {$Y$} } 
\ligne{
\larangea {} {} {\bfix B} {$T$} {\bfix W} 
\larangeb {\bfix W } {$Y$} {\bfix B} {\bfix W } } 
\ligne{
\larangea {} {} {\bfix B} {\bfix W} {\bfix W} 
\larangeb {$Y$} {\bfix B} {\bfix W } {\bfix B} } 
\vskip-6pt
\demitrait
\ligne{
\larangea {$1_6$} {\bfix W} {$Y$} {$Z$} {\bfix W} 
\larangeb {\bfix W} {\bfix W} {\bfix W} {\bfix W} }  
\ligne{
\larangea {} {} {\bfix B} {$Z$} {\bfix W} 
\larangeb {\bfix W} {\bfix W} {\bfix W} {\bfix W} }  
\ligne{
\larangea {} {} {\bfix B} {\bfix W} {\bfix W} 
\larangeb {\bfix W} {\bfix W} {\bfix W} {$Y$} }  
\vskip-6pt
\demitrait
\ligne{
\larangea {$1_7$} {\bfix W} {$Y$} {\bfix W} {\bfix W} 
\larangeb {\bfix W} {\bfix W} {\bfix W} {$X$} }  
\ligne{
\larangea {} {} {\bfix B} {\bfix W} {\bfix W} 
\larangeb {\bfix W} {\bfix W} {\bfix W} {$X$} }  
\ligne{
\larangea {} {} {\bfix B} {$Y$} {\bfix W} 
\larangeb {\bfix W} {\bfix W} {\bfix W} {\bfix W} }  
\vskip-6pt
\demitrait
}
\hfill}
}
\vskip 10pt
   Looking at each entry of the table attached to a cell, we can see that there is at
most a single configuration which is compatible with the application of a rule of~$A$.
In the other configurations, there is either a state~{\bf B} in a position where
the state~{\bf W} is expected or the converse situation.
Moreover, the admissible configuration occurs only for the cells which are on the yellow
line and never for the others. Accordingly, the rule which consists in applying the 
rule of~$A$ when there is one for that and to leave the current state unchanged otherwise
works more easily here. This completes the proof of Theorem~\ref{reffinment2} in the
case of the heptagrid.
\vskip 10pt
   Let us now look at the same problem in the case of the dodecagrid. This, time, we can 
take advantage of a bigger number of neighbours and of their spatial display to strengthen 
the difference between a cell of the yellow line which is implemented as indicated in
Subsection~\ref{line_dodec_n+1} and the cells which does not belong to this line. The 
way in which we establish this difference is illustrated by Figure~\ref{dodec_n}.
In this figure, the yellow colour represents the states of~$B$ which, by construction,
are in bijection with those of~$A$. As previously, the green colour is associated
with the state~{\bf W} which corresponds to the quiescent state~0 of~$A$, and the red 
colour is associated with the state~{\bf B} which corresponds to the state~1 of~$A$.
\vskip 10pt
\vtop{
\vspace{-10pt}
\setbox110=\hbox{\epsfig{file=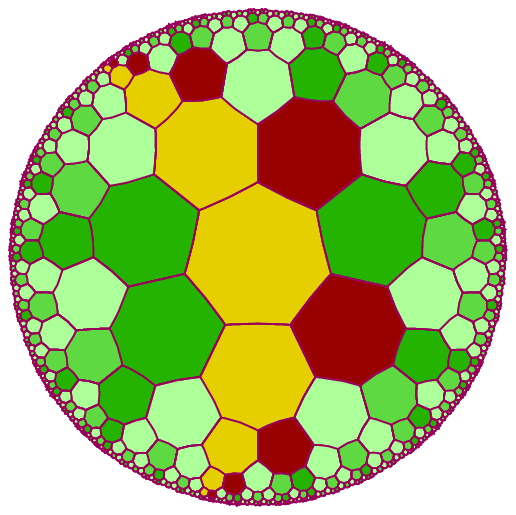,width=240pt}}
\ligne{\hfill
\PlacerEn {-125pt} {0pt} \box110
\hfill
}
\vspace{-5pt}
\begin{fig}
\label{hepta_n}
\leurre
Implementation of a cellular automaton in the heptagrid.
The yellow cells represent the line of tiles used for the $1D$ CA.
The green cells represent the cells which receive a particular state among the states
of the $1D$ CA.
\end{fig}
}
\vskip 10pt
   Now, each cell of the yellow line has four red neighbours. Numbering the cells
as indicated in Subsection~\ref{line_dodec_n+1}, the faces with a red neighbour are:
0, 3, 9 and~10, see figure~\ref{cut_dodec_n} which represents a cut in the plane
of the face~4 of a yellow cell. Due to the fact that face~0 is on the plane~$\Pi_0$, 
we can see only three red faces on the cells of the yellow line in Figure~\ref{dodec_n}.

   Now, let us consider the cells which do not belong to the yellow line. 
How many faces can they share with a red cell?
The answer to this question is: at most~2. 

   Consider any cell.
Its faces belong to planes which are either perpendicular or non-secant. If the faces have
a common vertex, then their supporting planes have a common line which supports an edge
of the faces. Now, if two faces~$F_0$ and~$F_1$ do not have a common vertex, there are 
two situations. In the first situation, there is a third face~$F_2$ such that $F_2$ 
shares a side~$s_0$ with~$F_0$ and another one~$s_1$ with~$F_1$. As $s_0$ and~$s_1$
do not have a common point, there is a side~$s_2$ of~$F_2$ which is in contact with 
both~$s_0$ and~$s_1$: $s_2$~is supported by a line which is the common perpendicular 
of the lines supporting~$s_0$ and~$s_1$. As the planes supporting the faces~$F_0$ and~$F_1$ 
are both perpendicular to~$F_2$, the line supporting~$s_2$ is also the common perpendicular
of the planes supporting~$F_0$ and~$F_1$. In the second situation, such a third face 
does not exist but then, faces~$F_0$ and~$F_1$ are opposite in the dodecahedron: they are
the reflection of each other under the reflection in the centre of the 
dodecahedron. Indeed, if we consider the face which is opposite to~$F_0$, the faces
which do not touch~$F_0$ and for which there is a third face sharing a common edge
with $F_0$ and its opposite face~$F_0'$ are all around the face~$F_0'$. And so,
if $F_1$ is none of them it is $F_0'$. Now, in this case, the perpendicular raised
from the centre of~$F_0$ is also the perpendicular raised from the centre of~$F_1$.
\vskip 10pt
\vtop{
\vspace{-10pt}
\setbox110=\hbox{\epsfig{file=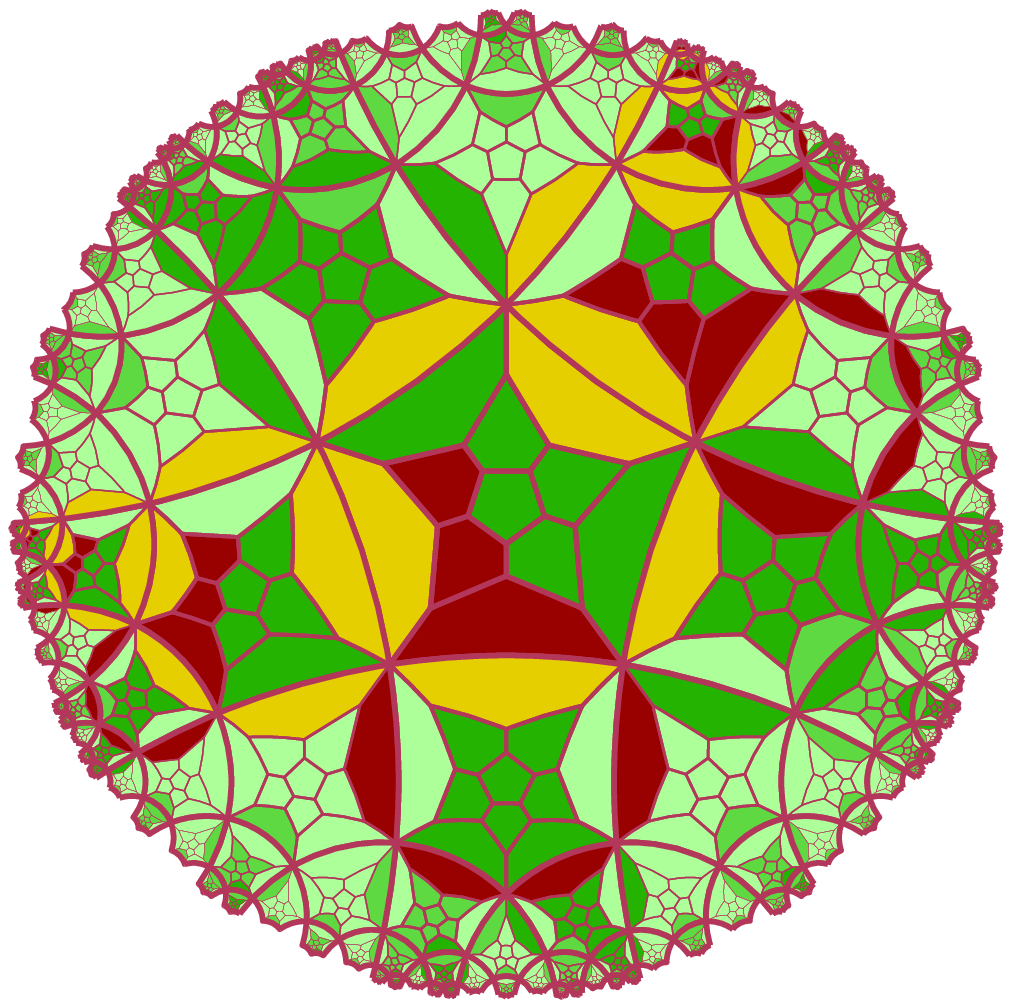,width=240pt}}
\ligne{\hfill
\PlacerEn {-125pt} {0pt} \box110
\hfill
}
\vspace{-5pt}
\begin{fig}
\label{dodec_n}
\leurre
Implementation of a cellular automaton in the heptagrid.
The yellow cells represent the line of tiles used for the $1D$ CA.
The green cells represent the cells which receive a particular state among the states
of the $1D$ CA.
\end{fig}
}
\vskip 5pt
From these geometrical considerations, we have that two neighbours of a cell have
no face in common. Now, consider two faces~$F_0$ and $F_1$ of a yellow cell~$C$
sharing an edge. We know that around this edge there are four dodecahedra: $C$ itself,
the reflection~$C_{F_0}$ of~$C$ in face~$F_0$, the reflection~$C_{F_1}$ of~$C$ in 
face~$F_1$ and a fourth one, $D$, which is the reflection of~$C$ in the common edge 
of $F_0$ and~$F_1$, see Figure~\ref{cut_dodec_n}. Now, as the planes supporting 
$F_0$ and~$F_1$ are perpendicular, this means that $D$ shares a face with~$C_{F_0}$ 
as well as another face with~$C_{F_1}$. Accordingly, there are neighbours of a red 
neighbour of a yellow cell which are in contact with two red neighbours. They are not 
in contact with other red cells. Indeed, in the above situation, consider that $C$~is 
a yellow cell and that $C_{F_0}$ and~$C_{F_1}$ are red ones. Then $D$~is green and it 
is not in contact with another red cell: the closest red cell is~$E$, the one which 
shares the face~3 of~$C$ with~$C$. Now, the plane~$\Pi$ of the face~3 of~$C$ cuts 
the space in two half-spaces: one contains~$E$, the other contains~$C$ and also 
$C_{F_0}$, $C_{F_1}$ and~$D$. Assuming that $C_{F_1}$ is in contact with~$\Pi$, if not 
it is the case of $C_{F_0}$, the above analysis on the faces of a dodecahedron tells us 
that the plane of the face shared in common by~$D$ and~$C_{F_1}$ is non-secant 
with~$\Pi$ and so, $D$ and~$E$ are far from each other. Note that a cell which does 
not belong to the yellow line, which is a neighbour of a cell of the yellow line and 
which is green is not the neighbour of a red cell not belonging to the yellow line. 
This comes from the same analysis as two neighbours of a cell are not neighbours of 
each other.

\vskip 10pt
\vtop{
\vspace{-10pt}
\setbox110=\hbox{\epsfig{file=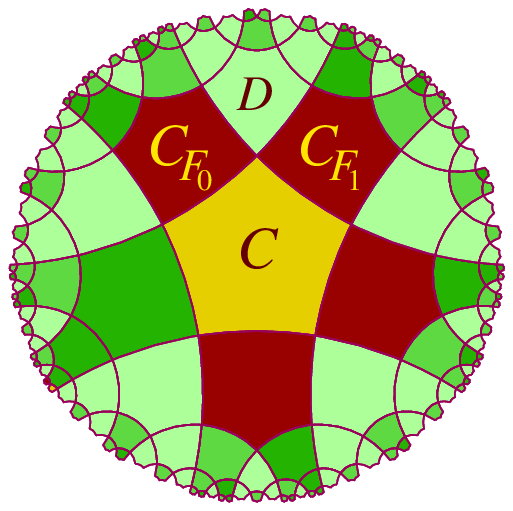,width=240pt}}
\ligne{\hfill
\PlacerEn {-125pt} {0pt} \box110
\hfill
}
\vspace{-5pt}
\begin{fig}
\label{cut_dodec_n}
\leurre
Implementation of a cellular automaton in the heptagrid.
The yellow cells represent the line of tiles used for the $1D$ CA.
The green cells represent the cells which receive a particular state among the states
of the $1D$ CA.
\end{fig}
}
\vskip 5pt

   From this analysis, it is clear that a yellow cell has four red neighbours exactly and
that all other cells have at most two red neighbours. In the case of four neighbours,
they are exactly those defined above and illustrated by Figures~\ref{dodec_n} 
and~\ref{cut_dodec_n} and the determination of which neighbours plays the role
of the left- or right-hand side neighbour is immediate. Accordingly, the rule which 
consists in applying the rule of~$A$ when there are four red neighbours and to leave the
current state unchanged when this not the case works still more easily here. This 
completes the proof of Theorem~\ref{reffinment2}.

   Now, we can see that from Theorem~\ref{reffinment2} we have as an immediate
corollary:

\begin{cor}\label{univ2_hepta_dodec}
There is a weakly universal rotation invariant cellular automaton on the heptagrid, as well
as in the dodecagrid with two states exactly.  
\end{cor}

   In both cases, we apply the construction defined in the proof of 
Theorem~\ref{reffinment2} to the elementary cellular automaton with rule~110.
Now, if we look at the transitions of rule~110, we can see that 0~is a quiescent state,
that it is fixed for the context~$1,0$ and that~1 is fixed for the context~$0,0$. This
proves that the elementary cellular automaton with rule~110 is fixable.
Consequently, applying Theorem~\ref{reffinment1} to this $1D$-cellular automaton,
we get:

\begin{cor}\label{univ2_penta}
There is a weakly universal rotation invariant cellular automaton on the pentagrid with two 
states exactly.
\end{cor}

\section{Conclusion}

   With this result, we reached the frontier between decidability and weak universality
for cellular automata in hyperbolic spaces: starting from 2~states there are weakly 
universal such cellular automata, with 1~state, there are none, which is trivial.

   What can be done further?

   In fact there are at least three possible directions. The first one is the frontier
between decidability and undecidability which requires the simulation of a cellular
automaton or of a Turing machine, in both cases, starting from a finite initial 
configuration. There is a result by Lindgren and Nordahl, see~\cite{linno90},
to the smallest universal Turing machines known at the present moment, 
see~\cite{mmCorkENTCS}, we obtain a deterministic $1D$-cellular automaton which is 
universal with 12~states. Now, this result cannot be immediately transported
to the tilings we have considered here as the general frame considered in Section~\ref{n+1}
as well as in Section~\ref{n} defines an initially infinite configuration.
Moreover, we have no result in 
the other direction, except the trivial case of a unique state. In particular, it is 
not known whether there is an analogue of Codd's theorem in the case of the hyperbolic 
plane. 

   The second direction starts with the remark that Corollaries~\ref{univ2_hepta_dodec}
and~\ref{univ2_penta} deal with only three tilings. Now, it is well known that in the
hyperbolic plane, there are infinitely many tilings on which we can implement cellular 
automata. And so, what can be said for these cases? 

   For what is the plane, the same technique as described in Section~\ref{n} works in
all tilings of the form $\{p,4\}$ and $\{p$+$2,3\}$ with $p\geq5$. The cases $p=5$ 
correspond to the pentagrid and the heptagrid respectively. Most probably, this work also
for $p=6$ and $p=7$ in the pentagrid. Now, starting from $p=7$ in the heptagrid and $p=9$
in the pentagrid, we can use a technique similar to
the one used for the dodecagrid: we can decide that the yellow cells have four red 
neighbours put at appropriate contiguous places around them and this will be enough to
distinguish yellow cells from the others.

    There is a third direction. The result proved in this paper suffers the same defect
as the result indicated in~\cite{mmarXiv3} with 3~states. The results proved in this 
paper can be obtained in a not too complicate manner by an appropriate implementation 
of rule~110 which is weakly universal, s already mentioned. In the case of the dodecagrid,
the author proved a similar result with 3~states but involving a much more elementary
construction which is also an actual $3D$~construction. In the case of the heptagrid, 
he obtained 4~states with an actual planar construction, see~\cite{mmarXiv4st,mmTCS4} and 
the best result known for the pentagrid is 9~states, see~\cite{mmsyPPL}, again with 
elementary tools and using an actual planar construction. What can be done in this 
direction is also an interesting question.

   Accordingly, there is some work ahead, probably the hardest as we are now so close 
to the goal.


\begin{thebibliography}{5}
\def\vvv{\vspace{-3pt}}
\bibitem{cook}
M.~Cook.
Universality in elementary cellular automata,
{\it Complex Systems}, (2004), {\bf 15}(1), 1-40.
\vvv
\bibitem {linno90}
Lindgren~K. and Nordahl~M.G.,
Universal computation in simple one-dimensional cellular automata. Complex
Systems, {\bf 4}, 299--318, (1990).

\vvv
\bibitem{mmbook1}
M. Margenstern,
Cellular Automata in Hyperbolic Spaces, Volume 1, Theory,
{\it OCP}, Philadelphia, (2007), 422p.

\vvv
\bibitem{mmbook2}
M. Margenstern,
Cellular Automata in Hyperbolic Spaces, Volume 2, Implementation and
computations,
{\it OCP}, Philadelphia, (2008), 360p.

\vvv
\bibitem{mmarXiv4st}
M. Margenstern,
A new universal cellular automaton on the ternary heptagrid,
{\it arXiv}:0903.2108[cs.FL], (2009), 35pp.

\vvv
\bibitem{mmCorkENTCS}
M.~Margenstern,
Surprising Areas in the Quest for Small Universal Devices,
{\it Electronic Notes in Theoretical Computer Science},
{\bf 225}, (2009), 201-220.

\vvv
\bibitem{mmarXiv3}
M. Margenstern,
A weakly universal cellular automaton in the hyperbolic 3D space with three states,
{\it arXiv}:1002.4290[cs.FL], (2010), 54pp.

\vvv
\bibitem{mmTCS4}
M. Margenstern,
A universal cellular automaton on the heptagrid of the hyperbolic plane with
four states,
{\it Theoretical Computer Science}, (2010), accepted.
 
\vvv
\bibitem{mmgsFI}
M. Margenstern, G. Skordev,
Tools for devising cellular automata in the hyperbolic 3D space,
{\it Fundamenta Informaticae}, {\bf 58}, {\rm N$^\circ$2}, (2003), 369-398.


\vvv
\bibitem{mmsyPPL}
M. Margenstern, Y. Song,
A new universal cellular automaton on the pentagrid,
{\it Parallel Processing Letters},
{\bf 19}(2), (2009), 227-246.


\vvv
\bibitem{wolfram}
S.~Wolfram.
A new kind of science, {\it Wolfram Media, Inc.}, (2002).


\end{thebibliography}
\end{document}